# Large-area nanoengineering of graphene corrugations for visible-frequency graphene plasmons

*Gergely Dobrik[1], Péter Nemes-Incze[1], Bruno Majerus[2], Péter Süle[1], Péter Vancsó[1], Gábor Piszter[1], Miklós Menyhárd[1], Benjámin Kalas[1], Péter Petrik[1], Luc Henrard[2]*

*& Levente Tapasztó[1*]*

1. Institute for Technical Physics and Materials Science, Centre for Energy Research, 1121 Budapest, Hungary * Email: tapaszto@mfa.kfki.hu

2. Department of Physics, Namur Institute of Structured Matter, University of Namur, Rue de Bruxelles 61, 5000, Namur, Belgium

**An effective way to engineer the properties of graphene is via quantum confinement of its carriers, most often realized through physical edges. However, edges come with major drawbacks, deteriorating mobility and strongly suppressing plasmon resonances. Here, we demonstrate a simple, large-area, edge-free nanostructuring technique, based on amplifying random nanoscale structural corrugations to a level where they efficiently confine carriers, without inducing significant inter-valley scattering. This soft confinement, allows the low-loss lateral ultra-confinement of graphene plasmons, scaling up their resonance frequency from native terahertz to commercially relevant visible range. Visible graphene plasmons localized into nanocorrugations mediate several orders of magnitude stronger light-matter interactions (Raman enhancement) than previously achieved with graphene, enabling the detection of specific molecules from femtomolar solutions or ambient air. Moreover, nanocorrugated graphene sheets also support propagating visible plasmon modes revealed by scanning near-field optical microscopy observation of their interference patterns.**

The nanoscale confinement of its charge carriers is an effective approach for engineering the properties of graphene, e.g., through opening sizeable band gaps[1], or tuning its plasmon resonance frequencies[2]. While transverse confinement – at the origin of the unique properties of 2D crystals – can easily be achieved by isolating single layers, an efficient in-plane confinement usually requires much harsher modifications, such as defining physical edges[1,3] or employing covalent chemistry[4,5]. Such hard potentials, acting through strong inter-valley scattering of charge carriers, significantly reduce charge carrier mobility[6], and strongly damp plasmon resonances[7]. While these adverse effects can be tolerated in larger structures, they become increasingly prohibitive in the sub-10 nm range [8]. Softer (smoother) potentials, avoiding inter-valley scattering, are usually not effective for confinement, since intra-valley backscattering is pseudospin-suppressed in graphene [9]. Nevertheless, this suppression can be lifted by pseudo-magnetic fields emerging from mechanical strain[10,11]. Edge-free confinement has previously been achieved in graphene by creating local *p-n* junction resonators with ~50 nm characteristic lateral size [12,13]. However, downscaling the confining structure size into the sub-10 nm range, as well as defining macroscopic arrays of such *p-n* resonators is highly challenging. Here, we achieved the sub-5 nm edge-free confinement of charge carriers by employing strong nanoscale corrugations. Random nanocorrugations are characteristic to 2D crystals, conferring them mechanical stability. However, the associated relatively weak distortions, only slightly perturb the atomic and electronic structure, and are primarily regarded as a source of disorder. By increasing their amplitude, one only expects to amplify the disorder. By contrast, we found that increasing the corrugation beyond a critical level (< 5 nm lateral size, > 0.4 aspect ratio) enables an efficient modification of the electronic structure, through nanoscale confinement of charge carriers, while largely avoiding their inter-valley scattering. This soft confinement mechanism extends the frontiers of confinement into the ultra-small limit, by eliminating the detrimental effects of edges and strong inter-valley scattering. Graphene plasmonics is a prominent field that can benefit from such soft nanostructuring. Graphene plasmons possess many intriguing properties, outperforming conventional metal nanoparticles, in terms of mode volume confinement, environmental stability and biocompatibility[14, 15]. However, exploiting graphene plasmons in the commercially relevant visible range is hindered by their significantly lower (THz) native resonance frequency. In the classical Drude approximation, the resonance frequency of graphene plasmons scales as $\omega_p \propto D^{-1/2}$, where D is the confining structure size[2]. Experiments employing arrays of graphene nanoribbons[7, 16] and quantum dots[17], confirmed that by decreasing the nanostructure size, the plasmon resonance frequency can indeed be increased. Graphene nanoribbons of widths below 100 nm, hosting IR plasmons have been employed to enhance the detection of various gas molecules with infra-red spectroscopy[18]. However, further decreasing the size of nanostructures, substantially increases losses that become prohibitive before plasmon frequencies could reach the visible range, estimated to happen for graphene nanostructures below 10 nm lateral size [2]. Charge carrier scattering on graphene edges was identified as one of the main reasons for damping graphene plasmon resonances[8,16]. Here, we show that the simple and scalable soft nanoengineering technique based on amplifying random nanoscale corrugations, allows the edge-free lateral confinement of graphene plasmons into ultra-small (sub-5nm) areas, scaling up their resonance frequency into the commercially relevant visible range. Visible frequency graphene plasmons, enable at least three orders of magnitude stronger Raman enhancements than previously achieved with graphene, allowing detection of molecules with well below part per trillion level sensitivity, and high selectivity. Furthermore, the interaction of nearby localized plasmons can give rise to propagating visible graphene plasmon modes in nanocorrugated graphene.

# 1. Results

## 1.1. Extremely nanocorrugated graphene sheets

Graphene sheets with unprecedentedly strong nanoscale corrugations have been prepared by cyclic thermal annealing (typically 2 - 4 cycles between RT and ~400°C) of mechanically exfoliated graphene flakes on standard $SiO_2$/Si substrates (see SI section I. for details). Fig. 1a,b show the nanoscale topography obtained from Scanning Tunneling Microscopy (STM) investigations of electrically contacted flakes, revealing a heavily nanocorrugated graphene structure.

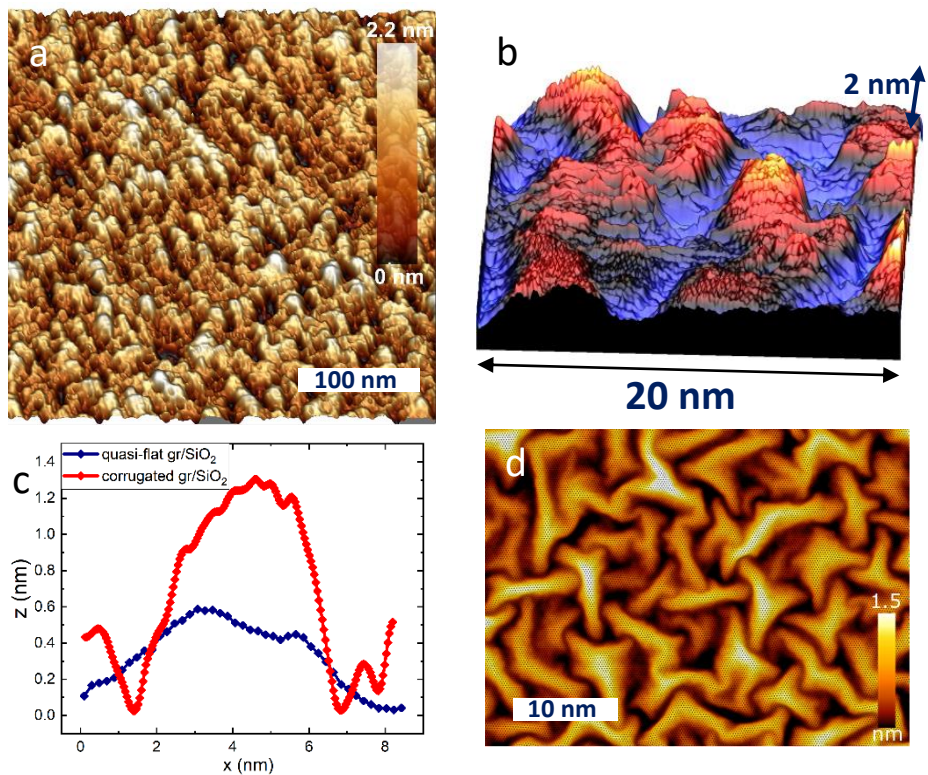

*Fig.1 Extremely nanocorrugated graphene sheets. a, b) Topographic STM images of graphene sheets prepared by cyclic thermal annealing, displaying particularly strong nanoscale corrugations, with lateral size below 10 nm, and nanometre height. c) Typical graphene ripple geometries measured for nanocorrugated ($h_{max}/R \sim 0.5$) and as-exfoliated (quasi-flat) graphene ($h_{max}/R \sim 0.15$) on $SiO_2$. d) Molecular Dynamics simulation of nanocorrugated graphene under compression of the Si substrate.*

The RMS roughness value extracted from topographic STM images is about 0.5 nm. This is almost double of the RMS value measured in graphene on $SiO_2$ (0.27 - 0.35 nm)[19]. The characteristic aspect ratios ($h_{max}/R$) of graphene nanocorrugations are typically of 0.4 – 0.5 (Fig. 1c). To ensure that the topography obtained by STM is not strongly influenced by the electronic structure, we also performed AFM investigations (SI Fig. 2). The results clearly highlight the extreme nature of the nanoscale deformations (corrugation) of nanobuckled graphene sheets, especially as scanning probe measurements tend to underestimate the aspect ratio of nanoscale features due to tip convolution effects[20]. The ability to create heavily nanocorrugated graphene sheets, originates from the interplay of two strong and antagonistic effects. Namely, the large strains emerging in supported graphene during thermal processing (owing to its negative thermal expansion coefficient[21]), and its strong adhesion to the $SiO_2$ substrate[22]. Although nanoscale deformations of graphene can be quite different from those of classical membranes[23], molecular dynamics simulations (see Methods for details) of graphene reproduce surprisingly well the characteristic nanoscale lateral size and high aspect ratio of the experimentally observed corrugations, upon compressing the Si substrate (Fig. 1d).

Raman spectroscopy investigations of heavily nanocorrugated graphene samples with 532nm excitation wavelength, reveal no significant disorder (D) peak (SI Fig. 3a), evidencing the absence of inter-valley scattering, in spite of the heavily deformed nanoscale structure. Further details of the Raman analysis are given in SI section III.

**1.2. Electronic structure of graphene nanocorrugations**

To investigate the influence of nanoscale deformations on the electronic structure of graphene, we have performed tunneling spectroscopy measurements on nanocorrugations (Fig. 2a).

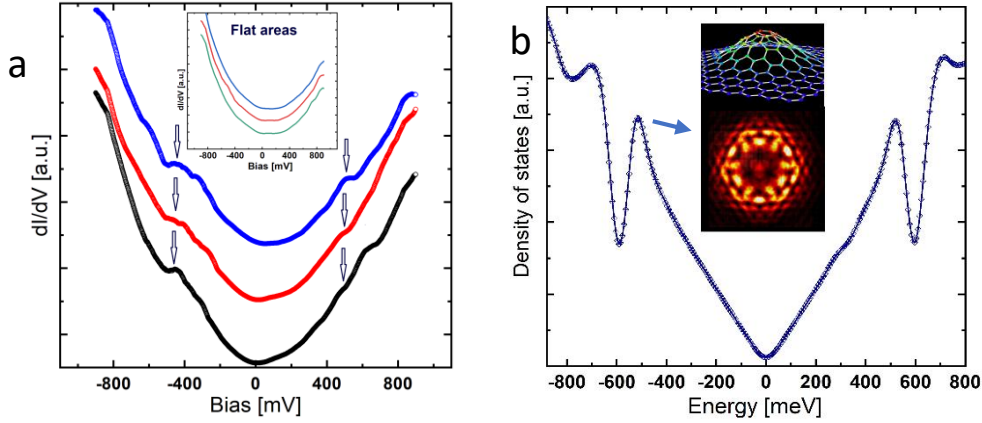

***Fig.2. Electronic structure of graphene nanocorrugations.*** *a) Tunneling spectra measured on D ~ 5 nm graphene corrugations with aspect ratios ($h_{max}/R$) ~ 0.4 - 0.5, displaying peak/shoulder-like features around +/- 450 mV. Inset shows tunneling spectra recorded on quasi-flat areas of the sample. b) DFT calculated electronic density of states averaged over a model graphene nanoprotrusion of similar aspect ratio ($h_{max}/R$ ~ 0.4). Inset shows the spatial distribution of the local density of states (LDOS), revealing the localization of specific electronic states on the nanocorrugation.*

Tunneling spectra acquired on nanocorrugations display distinctive features (peaks/shoulders) around +/- 450 *meV* compared to spectra measured on quasi-flat areas of the same sample. Similar peaks in the tunneling spectra have previously been identified as signatures of charge carrier confinement in one-dimensional sub-5nm graphene wrinkles[24]. From peak positions, one can estimate the characteristic confinement size to be of about 3 - 4 nm, in our case[24]. For further tunneling spectroscopy data, see SI section IV. To confirm our interpretation of the observed features in the tunneling spectra, we have performed DFT calculation on a model graphene corrugation of similar geometry (see Methods for details). The calculated density of states averaged over the graphene nanocorrugation (Fig. 2b) displays good agreement with the experimental tunneling spectra. We ensured that the resulting peaks do not originate from superlattice effect, but are characteristic to nanoscale graphene protrusions, by performing calculations with different supercell sizes. Plotting the spatial distribution of the calculated local density of states at energies near the LDOS peak, clearly evidences electronic states localized on the graphene nanoprotrusion (Fig. 2b inset), confirming the ability of graphene nanocorrugations with high aspect ratios to confine electronic states. The ability of structural deformations (mechanical strain) to spatially confine the charge carriers of graphene has already been demonstrated both theoretically[25,26] and experimentally[27,28]. The confinement is enabled by the large pseudo-magnetic fields emerging from nanoscale graphene corrugations (SI Fig. 5) that confines charge carriers by curving their classical trajectories into closed orbits around the peaks of the inhomogeneous pseudo-magnetic field[26,27].

**1.3. Huge Raman enhancement on nanocorrugated graphene sheets.**

An intriguing question is, how the above discussed modifications of the atomic and electronic structure can impart novel functionality to graphene. The edge-free nature of the confinement can allow localized plasmons to persist at such ultra-high confinement levels, required to scale up their resonance frequency into the visible. A key signature of plasmons, is their strong local electric field, enabling a particularly strong enhancement of light-matter interaction, leading to applications such as surface enhanced Raman scattering (SERS) or enhanced emission[15]. Graphene is already known to provide moderate Raman enhancements, often referred to as Graphene Enhanced Raman Scattering (GERS), which is due to the chemical enhancement mechanisms (charge transfer). GERS is characterized by enhancement factors of order of tens, as demonstrated for several molecules, among them copper phthalocyanine (CuPc), zinc phthalocyanine (ZnPc), or rhodamine 6G (Rh6G) [29,30,31]. Here, we have performed Raman spectroscopy investigations of such molecules on strongly nanocorrugated graphene sheets. The most striking findings were observed when measuring the Raman spectra of nanocorrugated graphene sheets with 633 nm excitation wavelength. Such measurements revealed a huge Raman signal, without subjecting the corrugated samples to any solution, only exposing them to laboratory air (Fig. 3b). Raman spectra displaying large additional peaks upon air exposure were highly reproducible with different corrugated graphene samples and substrates (SI Fig. 6), even fabricated at different laboratories, to exclude the possibility of a local furnace contamination. The huge Raman signal picked up form air (Fig. 3b) with nanocorrugated substrates, can be attributed to copper phthalocyanine (CuPc) molecules[30].

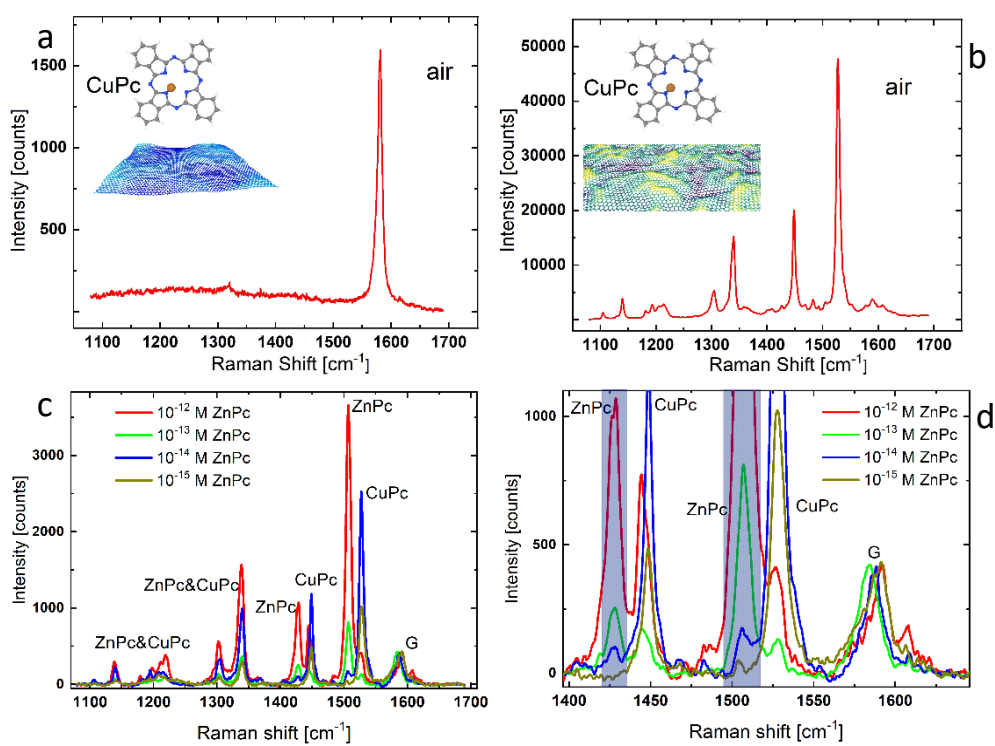

*Fig.3. Huge Raman enhancement on nanocorrugated graphene sheets. (a-b) Raman spectrum (633nm) of quasi-flat (a) and nanocorrugated (b) graphene sheets subjected to air, the latter detecting high intensity CuPc contamination peaks from nominally clean laboratory air. c) Raman signature of ZnPc molecules detected from down to femtomolar ($10^{-15}$ M) ZnPc solutions. CuPc peaks originate from air contamination. d) Zoom in of panel (c) on the region displaying distinct ZnPc (highlighted) and CuPc peak positions.*

While its concentration in the environment is unknown, CuPc is the synthetic dye molecule produced in the largest quantity in the world[32]. Similar to other byproducts of human activity[33, 34], CuPc seems to also be present practically everywhere. By AFM and STM investigations, we ensured that no unusual contamination can be observed on the surface of nanocorrugated graphene (SI section VII). Atomic resolution STM investigations revealed features that most likely correspond to CuPc molecules, present in a very low density (SI Fig. 7). Consequently, the observation of Raman peaks up to twenty times higher than the graphene G peak, must originate from an exceptionally strong enhancement mechanism, rather than a large quantity of adsorbed molecules. As expected, exposing quasi-flat graphene sheets to the same conditions does not result in any detectable CuPc signal (Fig. 3a). Based on the comparison (of G peaks) with quasi-flat graphene, we can estimate the lowest limit of the enhancement factor for CuPc on corrugated graphene to be of order of $10^5$. This is clearly beyond the range of chemical enhancement (below $10^2$ on flat graphene)[29,30], indicating plasmonic activity. The highly selective detection of phthalocyanine molecules from air is truly remarkable, and evidences the exceptional enhancement provided by nanocorrugated graphene for such molecules. However, the overall control of the process is quite limited. To overcome this, we have performed controlled experiments using zinc phthalocyanine (ZnPc) molecules. We prepared solutions with very low ZnPc concentrations by serial dilution of ZnPc solutions in isopropanol, then using the soaking technique[29], we have investigated the detection limit of nanocorrugated graphene substrates for ZnPc. The results are summarized in Fig. 3 c,d. Besides ZnPc Raman peaks, CuPc peaks can also be clearly detected, since Raman spectroscopy measurements were performed under ambient conditions with 633 nm excitation wavelength. We chose ZnPc as the test molecule, because two of its most intense Raman peaks are clearly distinguishable from those of CuPc (Fig. 3d). Our results reveal that ZnPc can be clearly detected from $10^{-14}$ M solutions, while its strongest peak is still detectable from $10^{-15}$ M solutions. This is a remarkable sensitivity, evidencing a three orders of magnitude lower detection than previously achieved with graphene ($10^{-11}$ M)[31]. We have also observed strong Raman enhancement on nanocorrugated graphene for a different type of molecule (R6G), at a different excitation wavelength (533 nm) (SI section VIII.). Various approaches for tuning the corrugation morphology and the corresponding plasmon resonance frequency are discussed in SI section IX.

## 2. Discussions

### 2.1. First-principles calculations of optical excitations in graphene nanocorrugations

The corrugation of graphene induces a nanoscale strain inhomogeneity that affects the chemical bonding and the local electronic structure as also evidenced by tunneling spectroscopy and DFT calculations (Fig. 2). In this case, the electromagnetic field enhancement should be modelled by *first-principles* approaches, to take into account such local effects. To maintain reasonable calculation times, we restrict our calculation to corrugated graphene with small unit cells (50 atoms – SI Fig 10). However, the main features reported for larger supercell are still captured (i.e. the localisation of the electronic states manifested in LDOS peaks, SI Fig. 11). Our calculations evidenced that nanoscale corrugations of high aspect ratio significantly modify the in-plane conductivity of graphene in the visible range compared to the constant value for flat sheets (SI Fig. 12). To understand the origin of these resonances, we have simulated the corresponding electron energy loss spectrum that reveals well defined peaks, associated with plasmon-like excitation in the visible range, in contrast to the featureless loss spectrum of flat graphene (Fig 4). We note that the maxima of the conductivity do not perfectly match the ones of the EELS spectrum, since EELS is more sensitive to collective excitations with high EM response, and dipolar selection rules break down in EELS[35]. We have also calculated and plotted the real space charge distributions corresponding to various peaks, some of them shown as insets of Fig. 4.

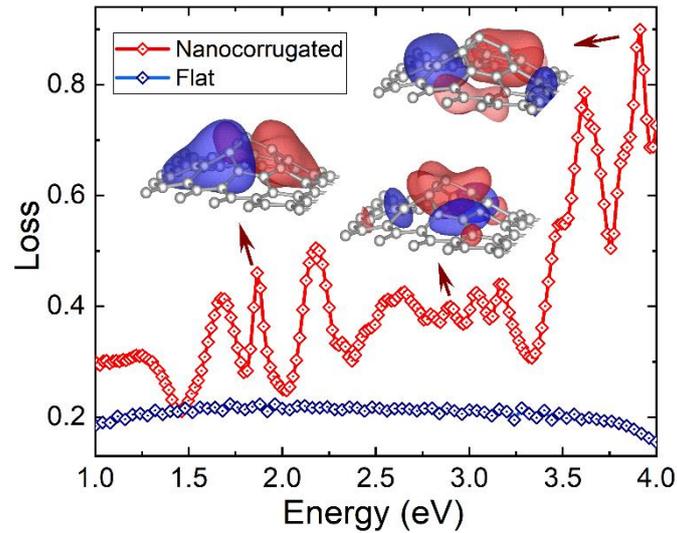

*Fig.4. Plasmonic excitations in graphene nanocorrugations.* *Calculated EELS spectrum of model graphene nanocorrugations, revealing several loss peaks in the visible range, in contrast to flat graphene. The insets display the charge distributions of optical excitations corresponding to loss peak near 1.9 eV, 2.9 eV, and 3.9 eV.*

Our computational results clearly confirm the ability of graphene nanocorrugations with high aspect ratio to host localized graphene plasmons of visible frequencies, supporting the plasmonic origin of the observed huge Raman enhancements in nanocorrugated graphene. A more quantitative modelling of the observed enhancements is highly challenging due to the strong influence of roughness and of the exact atomic scale geometry on the plasmon excitations and local electronic properties.

**2.2. Scanning near-field optical microscopy (SNOM) of nanocorrugated graphene**

SNOM measurements of nanocorrugated graphene samples, with 488 nm excitation wavelength (see Methods for details), reveal clear interference maxima in the proximity of edges, as well as fainter ~ 400 nm oscillations inward from edges and defects (Fig 5a). The observed patterns are highly similar to the SNOM images of plasmon interference patterns on doped quasi-flat graphene at IR frequencies[36,37]. However, in our case the interference patterns are observed at visible frequencies in nanocorrugated graphene, while quasi-flat graphene samples imaged under the very same conditions show no signs of plasmon interference (Fig. 5b).

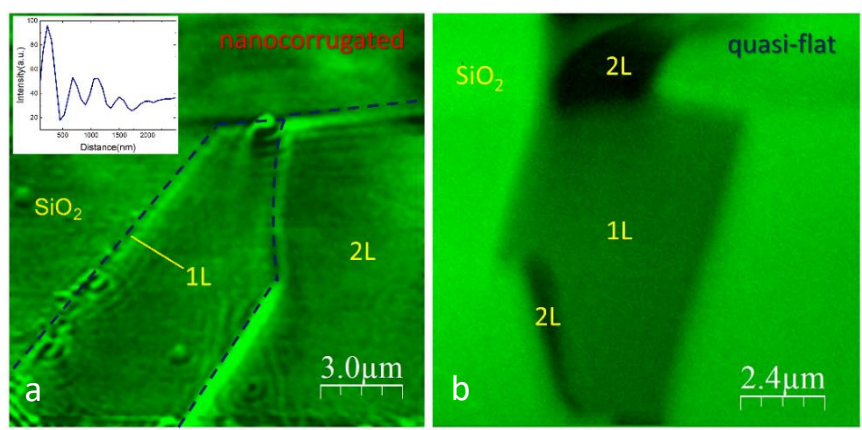

*Fig.5 Plasmon interference at visible frequency detected by SNOM. a) SNOM image ($\lambda$ = 488nm) of nanocorrugated graphene revealing clear interference maxima and oscillations in the proximity of edges (marked by dashed lines) and defects. Inset shows the line cut along the yellow line. b) SNOM image of quasi-flat graphene recorded under the same conditions, showing no interference patterns.*

Detecting interference patterns, clearly indicates the presence of propagating visible plasmons in nanocorrugated graphene. This is a surprising finding, since based on Raman spectroscopy and the proposed confinement mechanism, localized plasmons were primarily expected. However, localized and propagating plasmons are not mutually exclusive[38]. It has been theoretically predicted and experimentally confirmed that when separated nanostructures, hosting localized plasmons, are located in the close vicinity of each other ($< \lambda_{exc.}$), their interaction can give rise to propagating plasmon modes[39]. In nanocorrugated graphene the corrugations are typically located at least an order of magnitude closer than the excitation wavelength, enabling the interaction of localized plasmons to give rise to propagating modes, observed in our SNOM measurements (Fig.5a and SI Fig. 16). Furthermore, in such systems, the dispersion of transversal propagating modes crosses the light line, enabling the existence of plasmon wavelengths larger than the excitation wavelength. This can account for the oscillations of about 400 nm observed in SNOM images. Propagating modes with significantly smaller wavelength might also be present, but they cannot be resolved by our SNOM setup. Furthermore, propagating plasmon modes emerging form interaction of localized plasmons are predicted to persist even when disorder is present[38,40], which is clearly the experimental case. A more detailed understanding of visible graphene plasmons propagating in nanocorrugated samples requires further experimental and theoretical investigations. Confocal microscopy measurements with the same excitation wavelength show no interference patterns on nanocorrugated samples (SI Fig. 15b, 16), due to the inefficiency of coupling-in the freely propagating light (far-field) into graphene (due to large $q$ wave vector mismatch). By contrast, the near-fields emerging form the AFM tip of SNOM can excite propagating visible graphene plasmons at finite $q$, resulting in interference of the plasmons launched from the AFM tip and back-scattered from edges and defects.

We have also found signatures of plasmon activity in nanocorrugated graphene by EELS, spectroscopic ellipsometry and optical reflectance measurements. The results are presented in SI sections XII, XIII, and XIV, respectively. In the U.V. spectral range, the quenching of the π-plasmon in EELS measurements is well reproduced by DFT calculation and dielectric theory of EELS. In the visible range, ellipsometry and reflectance provide further evidence of the modification of the optical response of graphene related to the corrugations.

Regarding possible applications of visible graphene plasmons, SERS substrates based on nanocorrugated graphene offer a series of key advantages over conventional nanoparticle films, such as much simpler and cheaper fabrication, better reproducibility, and highly improved environmental stability, up to several months (SI Fig. 22). The high sensitivity allowing detection from air and femtomolar solutions, combined with high selectivity (detection of Pc molecules among all environmental contaminant species), highlights the potential of phthalocyanine molecules as efficient SERS labels. Furthermore, the fact that nanocorrugated graphene can host both localized and propagating visible graphene plasmons, evidences its unique potential as a versatile material platform for graphene plasmonics at visible frequencies.

**Methods**

*Raman spectroscopy and SNOM* measurements have been carried out using a Witec 300RSA+ confocal system, with 488, 532, and 633 nm excitation lasers. Single Raman spectra have been acquired with 0.5 mW laser power, and exposure times between 60 and 180 seconds, ensuring that the Raman signal did not change significantly during the measurement. SNOM measurements have been performed in reflection mode, with the SNOM sensor consisting of a Si cantilever and a hollow aluminium pyramid tip with 150 nm aperture.

*STM, AFM and tunneling spectroscopy* measurements were performed on a Nanoscope Multimode 8 setup operating under ambient conditions. Some STM and tunneling spectroscopy measurements have been performed in an RHK Pan Scan UHV setup. AFM measurements have been conducted in tapping mode. For STM measurements, the graphene flakes on insulating $SiO_2$/Si substrates were electrically contacted and identified under an optical microscope, enabling a guided landing of the Pt-Ir STM tip.

*MD simulations*. The corrugation of periodic 400x400 graphene (roughly 640k atoms) is obtained on Si(111) substrate by applying 0.75 GPa lateral pressure to the substrate at 300 K using Nose-Hoover npt MD simulation with the LAMMPS code[41]. (lcbop carbon potential[42], Lennard-Jones for Carbon-Si interaction: $\epsilon$=0.025, $r_{C\text{-}Si}$=3.3 Å, cutoff=6.0 Å, Tersoff-88 for Si[43], adaptive time stepping).

*Graphene nanocorrugation model geometries*. Gaussian nanowrinkles have been generated and subsequently optimized using the LAMMPS code (conjugated gradient minimization together with the relaxation of the simulation cell vectors). In order to avoid the collapse of the protrusions, a pulling force along the z-axis perpendicular to the protrusion's base sheet has been utilized. Using this approach, we were able to stabilize the dome height and relax the buckled structure to the possible minimum strain level.

*DFT band structure*. The fully periodic SIESTA code[44] has been used for fully self-consistent electronic structure calculations (double zeta+polarization (DZP) basis set, 400 Ry mesh cutoff, 15x15 k-mesh Monkhorst pack, PBE exchange-correlation functional). Diffuse function has also been added to the standard DZP basis set in order to improve the long range behavior of the wavefunction for high-quality *LDOS* maps. The SIESTA generated wavefunction file has been used for post-processing LDOS using the Denchar utility code.

*Optical calculations*. First principle simulations of the conductivity, EELS spectra and associated maps of the induced potential has been realized with GPAW code[45,46]. Local-density approximation (LDA)[47] for the exchange-correlation of electrons with a plane wave basis set with an energy cutoff of 400 eV was employed. The Brillouin zone was sampled with 32x32x1 Monkhorst-Pack mesh. For the calculation of the dielectric matrix we used an energy cutoff of 20 eV for the reciprocal lattice vectors **G** and **G'** and a smearing of 0.025 eV.

**Acknowledgements**

This work has been supported by the NanoFab2D ERC Starting Grant and the Graphene Flagship, H2020 GrapheneCore3 project no. 881603. L.T. acknowledges support from the NKFIH OTKA grant K 132869, and the Élvonal grant KKP 138144. P.N-I acknowledges the support of the "Lendület" program of the Hungarian Academy of Sciences, LP2017-9/2017. LH and BM acknowledge the use of the computational resources provided by the Consortium des Équipements de Calcul Intensif (CÉCI), funded by the Fonds de la Recherche Scientifique de Belgique (F.R.S.-FNRS) under Grant No. 2.5020.11 and by the Walloon Region and the support of the ARC research project N° 19/24-102 SURFASCOPE. G.D. and P.V. acknowledges support from the Bolyai fellowship of the Hungarian Academy of Sciences. The authors acknowledge valuable discussions with Philippe Lambin and Javier Garcia de Abajo.

**Author contributions**

L.T. conceived and designed the experiments. G.D. and P.N-I. performed the sample preparation, Raman and SNOM measurements. G.D performed the STM, AFM and tunneling spectroscopy investigations. P.S. performed Molecular Dynamics, geometry optimization and DFT LDOS calculations. B.M., P.V. and L.H. provided the optical calculations. P.P and B.K. performed the spectroscopic ellipsometry investigations. M.M. measured the EELS spectra. G.P. and G.D. conducted the reflectance spectroscopy measurements. L.T. wrote the paper. All authors discussed the results and commented on the manuscript.